*A 2572 and HCG 94 – galaxy clusters but not as we know them.*

*An X-ray case study of optical misclassifications*

H. Ebeling[1], C. Mendes de Oliveira[2,3] & D.A. White[1]

[1] *Institute of Astronomy, Madingley Road, Cambridge CB3 0HA, UK*
[2] *European Southern Observatory, Casilla 19001, Santiago, 19, Chile*
[3] *Instituto Astronômico e Geofisico, USP, C.P. 9638, São Paulo, 01065-970, Brazil*

**ABSTRACT**
We present the results of a spectral-imaging analysis of X-ray data obtained with the Position Sensitive Proportional Counter aboard the ROSAT Observatory in a 32 ks pointed observation of Hickson's Compact Group (HCG) # 94. Besides HCG 94, A 2572, a richness class 0 Abell cluster, is also contained in the central region of the field of view.

Both systems are at a redshift of $z \sim 0.04$ and are falling toward each other at a velocity of about 1000 km s$^{-1}$. Their three-dimensional spatial separation is probably of the order of an Abell radius; however, as yet, no clear signs of dynamical interaction are discernible in the X-ray.

We find HCG 94's gas temperature and unabsorbed X-ray luminosity to be far too high for a galaxy group thereby confirming the claim of Ebeling, Voges & Böhringer (1994) that HCG 94 should be classified as a galaxy cluster. The opposite is true for the Abell cluster A 2572, the optical richness of which has been overestimated due to the inclusion of HCG 94. In the X-ray, A 2572 appears at first sight like the prototypical binary cluster with two equally massive and X-ray bright subclusters in the process of merging. However, the available X-ray, optical, and radio data strongly suggest that A 2572 proper is in fact merely a loose group of galaxies, while the second component is a much richer and more distant cluster seen in superposition.

A deprojection analysis shows HCG 94 to host a moderate cooling flow; this picture is supported by a radial increase in the column density of absorbing material and a decrease in the gas temperature toward the cluster centre. HCG 94's total gravitating mass is much higher than what could be anticipated from its appearance in the optical. Our findings hence underline the need for X-ray selected cluster samples.

For all three clusters studied in this paper we find the baryon fraction to rise with radius and reach values of 15 to 30 per cent at the outer edge of our study regions. If any of these values is to be taken as representative of the overall baryon fraction of the Universe, then this result requires the latter to be open with $\Omega_0 < 0.35$ if a conflict with the baryon density derived from nucleosynthesis calculations is to be avoided.

**Key words:** X-rays: galaxies – galaxies: clusters: general – galaxies: clusters: individual: A 2572 – galaxies: clusters: individual: HCG 94 – cosmology: observations

## 1 INTRODUCTION

Clusters of galaxies are the most massive systems in the Universe and, as such, believed to form last in a hierarchical picture of structure formation. Accordingly, rich clusters grow through successive mergers of groups and poor clusters of galaxies, i.e. by a process that should be observable today. Indeed, a large fraction of clusters of galaxies are found to show substructure, both in the optical (e.g., Geller & Beers 1982; Dressler & Shectman 1988; Zabludoff, Huchra & Geller 1990; Rhee, van Haarlem & Katgert 1991) and in the X-ray (e.g., Forman et al. 1981; Jones & Forman 1984; Briel et al. 1991; White, Briel & Henry 1993; Forman & Jones 1994; Henry & Briel 1995), much of which can be explained naturally as being caused by ongoing or recent merging events.

The system A 2572/HCG 94 is situated at the north-western end of a supercluster which is also comprised of A 2589, A 2593, and A 2657. It is listed as number 22 in the supercluster catalogue of Postman, Huchra & Geller (1992); in the compilation of Zucca et al. (1993) it is referred to as SC 41/2. The core of the complex consisting of A 2572, A 2589, and A 2593 has a true spatial galaxy overdensity over its environment of more than 40 according to Zucca et al. and is itself again part of the even larger Perseus-Pegasus filament. Accordingly, the overall system can be expected to provide a dynamical environment for cluster evolution through merging.

The potential merger of A 2572 with HCG 94 was discovered serendipitously by Ebeling et al. (1994) when they searched the ROSAT All-Sky Survey (RASS, Voges 1992) data for X-ray emission from Hickson's Compact Groups of galaxies (HCGs, Hickson 1982, Hickson 1993). From the short observation of HCG 94 in the RASS ($t_{exp} = 466$ s) Ebeling and co-workers had already concluded that HCG 94 (also known as ARP 170) is not an isolated group of galaxies but rather the compact core of a proper cluster with a total X-ray luminosity in the $0.1 - 2.4$ keV band of about $1 \times 10^{44}$ erg s$^{-1}$. HCG 94 is thus brighter in X-rays than its close neighbour A 2572, a richness class 0 galaxy cluster (Abell, Corwin & Olowin 1989).

The two systems are only 17 arcmin apart and both at a redshift of $z \sim 0.04$. The literature redshift of A 2572, however, is based on spectroscopic data for only four galaxies, and more redshifts are needed for a detailed investigation into the cluster's spatial galaxy distribution (see Section 3 for a discussion of the optical properties of A 2572/HCG 94). A possible association between HCG 94 and A 2572 has also been pointed out by Pildis, Bregman & Evrard (1995) who find HCG 94 as a serendipitous source (30$'$ off axis) in their ROSAT pointing on HCG 93.

Although optical investigations show that a considerable fraction of HCGs are in some way related to more massive entities in their immediate environment, it is only in X-ray observations that the mass distribution of the overall system and, possibly, the interaction between its components can be observed directly. The deep ROSAT Position Sensitive Proportional Counter (PSPC) observation of A 2572/HCG 94 discussed in the following demonstrates once more the importance and power of X-ray observations for our understanding of cluster properties.

An Einstein-de Sitter cosmology with $h_0 \equiv H_0/100 = 0.5$ and $q_0 = 0.5$ is assumed throughout this paper.

## 2 OBSERVATIONAL PARAMETERS

HCG 94 was observed with the PSPC on December 4, 1993 for a total of 31560 s. The pointing was centred on HCG 94 at $23^h17^m16.8^s, +18°43'12''$ (equinox J2000) but includes A 2572 which is 17 arcmin to the east and just inside the inner ring of the PSPC window support structure.

The PSPC's circular field of view has a diameter of 2 degrees; the energy range covered by the XRT/PSPC configuration is $0.1 - 2.4$ keV. However, we limit the spectral analysis to the range from 0.2 to 2.0 keV where the detector response is least affected by calibration uncertainties. For imaging purposes photons of energies down to 0.1 keV are considered. The column density of Galactic Hydrogen in the direction of the telescope pointing is $(4.74 \pm 0.38) \times 10^{20}$ cm$^{-2}$ (Stark et al. 1992). The quoted error is the standard deviation of the values found at the neighbouring grid points.

In all of the following we restrict ourselves to those parts of the accepted observation time intervals where the 'Master Veto Rate' (parameter EE_MV in the SASS FITS data) and the error in the aspect solution (parameter ASP_ERR) remain below 150 cts s$^{-1}$ and 1 arcsec, respectively. We find no evidence for afterpulse contamination in the background spectrum. The total deadtime corrected exposure for our highest quality data is 25780 s.

## 3 X-RAY-OPTICAL-RADIO COMPARISON

Figure 1 shows the surface brightness distribution in the 0.1 to 2 keV range in a $50 \times 50$ arcmin$^2$ field centred on $23^h17^m48^s, +18°43'24''$. The raw surface brightness distribution has been accumulated from five images for energy bands which subdivide the broad 0.1 to 2 keV range of the PSPC. Each spectral subimage takes into account the energy dependence of the PSPC exposure map by using the energy dependent instrument maps constructed by S.L. Snowden and co-workers (Snowden et al. 1994). The image shown in Fig. 1 has been smoothed adaptively over areas containing a minimum number of counts $n_{min}^{asmooth} = 50$. By averaging over circular regions containing a fixed number of photons rather than over a circle of fixed radius, adaptive smoothing heavily suppresses background noise while at the same time maintaining the spatial resolution of the raw data in regions where the counts per pixel reach or exceed $n_{min}^{asmooth}$. The algorithm has been described in more detail by Rangarajan et al. (1995) who used it in their analysis of PSPC and HRI observations of the bright elliptical M86 in the Virgo cluster.

Figure 1 shows extended emission from the intracluster medium (ICM) around both HCG 94 and A 2572 out to radii of 0.9 and 0.6 Mpc, respectively (a redshift of 0.04 has been assumed in the conversion from angular to metric scale). The level of emission between the two clusters is significantly enhanced over the background on a scale of almost 2 Mpc in the NS direction. However, we find no obvious signs of violent interaction, and in particular no shock front, in the interface region. The emission from HCG 94 is elongated toward a position angle (p.a.) of about 30°; neither the p.a. nor the ellipticity of about 0.2 change noticeably within the inner 15 arcmin. The core of A 2572 is found to be double consisting of two X-ray peaks of similar intensity separated by about 3 arcmin which translates into 200 kpc if a redshift of $z \sim 0.04$ is assumed for both components. Again, the ICM between the two X-ray peaks appears undisturbed.

In addition to the extended emission from the two clusters, a number of point sources are detected, the most prominent ones being the two west of A 2572 and north-east of HCG 94, respectively. The true scale of the three main peaks from Fig. 1 and also the ratio of the clusters' brightness to that of the point sources become more apparent from Fig. 2 which shows a three-dimensional representation of the surface brightness distribution in the central area of Fig. 1 when viewed from the north. The scale is linear from $4.6 \times 10^{-4}$ to a maximum value of $5.1 \times 10^{-2}$ cts s$^{-1}$ arcmin$^{-2}$. Note the two brightest point sources next to A 2572 and HCG 94 and also the other compact sources that stand out more clearly in this representation than in Fig. 1.

The cores of the two main bodies of emission around A 2572 and HCG 94 are depicted in Figs. 3 and 4. We show the raw data binned into 10 arcsec pixels side by side with the adaptively smoothed images obtained from them. The signal-to-noise ratio is greater than 7 for all pixels in the adaptively smoothed hard band image (0.5 to 2 keV); for the soft band (0.11 to 0.41 keV) the respective value is 5. Note that all of the features detailed in the following are discernible in both the raw and the smoothed images alike.

- A profound lack of soft X-rays from the south-eastern component of A 2572, most notable in the very core, when compared to its northern neighbour.
- A significant ($> 2\sigma$) drop in surface brightness in the very centre of the southern X-ray peak in A 2572 in the hard band which leaves the remaining emission horseshoe-shaped.
- Diffuse, soft excess emission between the two main X-ray peaks in A 2572.

**Figure 1.** Exposure corrected image of HCG 94 and A 2572 in the PSPC's 0.1 to 2.0 keV band. The raw image (pixel size: 15″) has been smoothed adaptively using a circular top hat filter which encompasses a minimum of $n_{\min}^{\mathrm{asmooth}} = 50$ counts within the kernel; the maximum number of counts per pixel in the unsmoothed image is 80. The brightness scale is logarithmic and spans a range from $8 \times 10^{-4}$ to $5.3 \times 10^{-2}$ cts s$^{-1}$arcmin$^{-2}$.

**Figure 2.** Three-dimensional representation of the adaptively smoothed surface brightness distribution as depicted in Fig 1, viewed from the north. Shown is an area of radius 25′ centred on $23^h17^m48^s, +18°43'24''$. The scaling is linear and spans a range from $4.6 \times 10^{-4}$ to $5.1 \times 10^{-2}$ cts s$^{-1}$arcmin$^{-2}$.

**Figure 3.** The surface brightness distribution in the central 8 arcmin around the double core of A 2572. In the top two panels the emission in the hard energy band from 0.52 to 2.0 keV only is considered; in the lower two the energy range is from 0.11 to 0.41 keV. In each row the left-hand image shows the raw data binned into 10 arcsec pixels; the image to the right shows the same data after adaptive smoothing with $n_{\min}^{\mathrm{asmooth}} = 50$ and $n_{\min}^{\mathrm{asmooth}} = 25$ for the hard and soft energy bands, respectively. The maximum number of counts per pixel in the raw data is 26 in the hard band and 8 in the soft band.

- A north-south elongation in the emission from the core of HCG 94, possible due to the superposition of two sources. These are only marginally resolved in the hard band, but stand out more clearly in the soft X-rays (if at a somewhat different p.a.).

As far as its optical properties are concerned, HCG 94 is by no means representative of compact groups in general. It is one of the HCGs with the largest number of members, seven, six of which are either S0 or elliptical galaxies (Hickson 1982). Two of the seven are, however, fainter than $m_1 + 3$, the nominal magnitude limit in the definition of HCGs. The three central galaxies of the group are embedded in a bright envelope of light which is displaced to the eastern side of the cluster (in the direction of A 2572). HCG 94 also contains a large number of faint galaxies that may be members of the cluster. Redshifts for all seven group members were measured by Hickson et al. (1992) and were found to be accordant. Their median radial velocity is 12250 km s$^{-1}$ ($z = 0.0417$) with a dispersion of 479 km s$^{-1}$, the second largest value of all HCGs.

According to Sulentic (1987) and Palumbo et al. (1995), HCG 94 is among the Hickson Compact Groups with the highest galaxy-density environment. However, the galaxy counts these studies are based upon were obtained within radii of 30 arcmin or more and are hence contaminated by the galaxies around A 2572. Excluding the group itself, Rood and Williams (1989) counted galaxies out to 10 angular group radii which, for HCG 94, corresponds to about 14 arcmin, i.e. less than the angular separation of HCG 94 and A 2572. Within that area they find only three additional galaxies that are comparable in brightness to the group's member galaxies. This result is in conflict with the number of 12 galaxies that Kindl (1990, Table 8) found in the same area (and in the same magnitude range, i.e. at $m_B \lesssim 18$ but excluding obvious foreground galaxies). In an attempt to resolve the confusion, we counted galaxies within a third circle of 10 arcmin radius (corresponding to about 700 kpc, cf. region $A_2 \cup A_3$ in Fig.8) around HCG 94, and, again excluding the group proper, found four. This agrees with Kindl who also counts four galaxies within the same 10 arcmin radius, and suggests that the counts presented by Rood & Williams may have been obtained at brighter magnitudes.

**Figure 4.** The surface brightness distribution in the central 8 arcmin around the core of HCG 94 In the top two panels the emission in the hard energy band from 0.52 to 2.0 keV only is considered; in the lower two the energy range is from 0.11 to 0.41 keV. In each row the left-hand image shows the raw data binned into 10 arcsec pixels; the image to the right shows the same data after adaptive smoothing with $n_{\min}^{\mathrm{asmooth}} = 50$ and $n_{\min}^{\mathrm{asmooth}} = 25$ for the hard and soft energy bands, respectively. The maximum number of counts per pixel in the raw data is 36 in the hard band and 8 in the soft band.

We have obtained spectra for 11 galaxies in HCG 94 and its neighbourhood, five of which did not have measured redshifts previously. None of the galaxies show emission lines in their spectra. For one of the galaxies (HCG 94b) with a good S/N spectrum the internal velocity dispersion was determined. It is found to be normal for HCG 94b's luminosity and effective radius and places the galaxy in a normal place on the fundamental plane (Mendes de Oliveira et al. 1994). The NIR colors for five of the galaxies in HCG 94 have also been measured and were found to be typical for the galaxies' luminosities (Mendes de Oliveira, Mackenty & Hickson 1995).

A 2572 has been classified as a richness class 0 cluster by Abell (1958). Abell, Corwin & Olowin (1989) list it as a BM type III system at $z = 0.0395$; the number of galaxies is given as 32 above the background. Since, at that redshift, an Abell radius ($1r_A = 1.5 h_0^{-1}$ Mpc) equals more than 40 arcmin, Abell's galaxy counts for A 2572 include HCG 94; without the latter A 2572 would have been too optically poor to be included in Abell's catalogue. According to Struble & Rood (1987) the redshift of 0.0395 for A 2572 is based on four galaxies. These four are, however, distributed over a considerable area, and two of them are in fact the two dominating ellipticals in HCG 94, NGC 7578A and B (Chincarini & Rood 1972).

Table 1 lists the positions and radial velocities of 19 galaxies in the field of Fig. 1. Where the coordinates in the literature were found to have errors in excess of 30 arcsec, we remeasured the galaxy positions from the POSS E plate using an astrometry solution based on the plate positions of the bright stars HD 219780, HD 219419, and HD 219499. The coordinates listed in Table 1 are accurate to about 10 arcsec. We also include the positions and radial velocities of four galaxies which are outside the field of Fig. 1 but lie within one Abell radius of either cluster centre and feature redshifts consistent with the cluster mean. For 12 of these 23 the redshifts have been extracted from the literature. Five new redshifts have been measured by one of us (CMdO) with the ESO 1.5m telescope as part of a project investigating the optical properties of optically poor clusters detected in the ROSAT All-Sky Survey. The positions of all 19 galaxies within the field of Fig. 1 and with redshift information are shown in Fig. 5.

Two conflicting redshifts have been found in the literature for galaxy 'r', and we list both of them in Table 1. According to Chincarini & Rood (1972) the galaxy is a cluster member, whereas the redshift measurement of Owen, Ledlow & Keel (1995) places it firmly in the background at $z = 0.1547$. Given the considerable uncertainty in the galaxy position and the limited spectral coverage of the instrumental setup used by Chincarini & Rood on the one hand, and the high quality of the spectrum obtained by Owen and co-workers on the other hand (M. Ledlow, private communication), we accept the higher of the two values as the true redshift of galaxy 'r'.

Discarding the foreground object UGC 12471 as well as the background galaxy 'r' and assigning all galaxies east of $23^h17^m45^s$

**Figure 5.** Positions of the 19 galaxies with measured redshifts superposed onto the hard-band X-ray contours [cf. Fig. 6] in the field of Fig. 1. See Table 1 for details.

| galaxy # | name | $\alpha$ (J2000) | $\delta$ | $v$ km s$^{-1}$ | $\Delta v$ km s$^{-1}$ | $v_{\rm lit}$ km s$^{-1}$ | $\Delta v_{\rm lit}$ km s$^{-1}$ | reference |
|---|---|---|---|---|---|---|---|---|
| a | UGC 12471 | 23 16 28.4 | +18 33 46 | | | 2063 | 5 | GH |
| b | NGC 7572 | 23 16 50.4 | +18 29 03 | 13084 | 36 | 13070 | | GH |
| c | | 23 16 58.1 | +18 45 55 | 12590 | 38 | | | |
| d | | 23 17 07.1 | +18 39 59 | 11600 | 120 | | | |
| e | | 23 17 08.9 | +18 39 46 | 11847 | 71 | | | |
| f | NGC 7578A, HCG 94b | 23 17 12.0 | +18 42 03 | 11907 | 66 | 11974 | 37 | HMHP |
| g | NGC 7578B, HCG 94a | 23 17 13.5 | +18 42 28 | 12187 | 108 | 12040 | 42 | HMHP |
| h | HCG 94d | 23 17 15.2 | +18 42 43 | 13132 | 72 | 13009 | 42 | HMHP |
| i | HCG 94e | 23 17 15.5 | +18 43 36 | | | 12250 | 103 | HMHP |
| j | HCG 94f | 23 17 18.6 | +18 44 21 | | | 12920 | 108 | HMHP |
| k | HCG 94g | 23 17 20.0 | +18 44 56 | 13497 | 80 | 13200 | 114 | HMHP |
| l | HCG 94c | 23 17 20.3 | +18 44 04 | 12043 | 54 | 12120 | 52 | HMHP |
| m | CGCG 454-025 | 23 17 21.1 | +18 03 22 | | | 11503 | | GH |
| n | CGCG 454-023 | 23 17 21.5 | +18 25 24 | | | 13237 | | GH |
| o | | 23 17 30.8 | +18 32 22 | 11721 | 44 | | | |
| p | NGC 7588 | 23 17 57.8 | +18 45 09 | 10850 | 35 | | | |
| q | NGC 7597 | 23 18 30.3 | +18 41 20 | | | 11048 | 44 | RC3 |
| r | | 23 18 39.2 | +18 41 23 | | | 42827 | 100 | OLK |
| | | | | | also: | 10838 | 230 | CR |
| s | NGC 7602 | 23 18 43.5 | +18 41 54 | | | 11610 | 71 | CR,RC3 |
| t | CGCG 454-036 | 23 19 21.3 | +19 01 13 | | | 12070 | | GH93 |
| u | I5312 | 23 20 29.6 | +19 19 27 | | | 11020 | 30 | BFHJG |
| v | CGCG 454-045 | 23 20 48.0 | +18 54 20 | | | 11546 | 37 | BFHJG |
| w | CGCG 454-049 | 23 21 25.1 | +18 31 27 | | | 11956 | | GH93 |

GH = Giovanelli R. & Haynes M.P., 1993, AJ, 105, 1271
HMHP = Hickson P., Mendes de Oliveira C., Huchra J.P., Palumbo G.G.C., 1992, ApJ, 399, 353
OLK = Owen F.N., Ledlow M.J., Keel W.C., 1995, AJ, 109, 14
RC3 = de Vaucoulers G., de Vaucouleurs A., Corwin H.G, et al., Third Reference Catalogue of Bright Galaxies, Springer 1991
CR = Chincarini G & Rood H.J., 1972, AJ, 77, 4
BFHJG = Beers T.C., Forman W., Huchra J.P., Jones C., Gebhardt K., 1991, AJ, 102, 1581

**Table 1.** Galaxies with measured redshifts in the field of Fig. 1. The galaxies m, u and v are actually outside that field but lie within one Abell radius of either of the two clusters and have radial velocities consistent with the cluster mean. Their positions may be uncertain by more than 10 arcsec.

to A 2572 and those west of the same delimiter to HCG 94, we arrive at the following redshifts and velocity dispersions for the two systems: $z = 0.04218$, $\sigma = 663$ km s$^{-1}$ (HCG 94, 14 galaxy redshifts); $z = 0.03893$, $\sigma = 480$ km s$^{-1}$ (A 2572, 7 galaxy redshifts). Due to the overlap of the galaxy populations of the eastern and western components these velocity dispersions should, however, be taken rather as upper limits; the velocity dispersion of the seven galaxies constituting HCG 94 proper is indeed only 479 km s$^{-1}$ (Hickson 1993).

Some galaxies within the field of Fig. 1 have also been observed at radio wavelengths. Galaxy 'r' (cf. Table 1 and Fig. 5) is a twin-tail radio source with an integrated flux density of 0.29 Jy and a core flux density of 15 mJy at 1.4 GHz (O'Dea & Owen 1985). The jets are bent backwards in an east-south-easterly direction in a way which O'Dea & Owen describe as one of the "best examples of the type of morphology expected from a galaxy with a turbulent galactic wake". Note, however, that this galaxy is at $z = 0.1547$ and thus *not* a member of the cluster A 2572. It's radio morphology is nonetheless typical of a galaxy falling toward a nearby cluster, which is why we take these radio data as a first indication for the presence of second cluster of galaxies at a redshift of $z \sim 0.155$ which is superimposed onto A 2572. Further, supporting evidence for the hypothesis that the X-ray peak north of A 2572 should in fact be attributed to this background cluster is presented in the following sections and summarized in Section 6.

To identify the sources and structures detected in Figs. 1 to 4 we superimpose the X-ray surface brightness contours onto optical images of the same area of the sky. Figure 6 shows the contours for a surface brightness map adaptively smoothed with $n_{\rm min}^{\rm asmooth} = 50$ in the hard energy band (0.52 to 2.0 keV) overlaid onto the POSS E image[*]. A blow-up of the central regions around HCG 94 and A 2572 is shown in Fig. 7. From Figs. 2, 3, 4, 6, and 7 a number of X-ray sources and features can be readily associated with optical counterparts.

• The south-eastern X-ray peak in A 2572 is centred on the bright elliptical NGC 7597 which has a measured redshift of $z = 0.0376$ (galaxy 'q' in Table 1 and Fig. 5). It has been classified as a cD and is possibly the result of a recent merger (Schombert 1987).

• The absorption feature in the very core of the emission around NGC 7597 coincides precisely with the position of the galaxy itself [see Figs. 3 and 7 (bottom)].

• The north-western X-ray peak in A 2572 is centred on a group of at least three galaxies all of which are much fainter than NGC 7597. None of them has a measured redshift.

• The soft point source seen just east (about 1 arcmin) of the

---

[*] The optical image is based on a digitized scan of the corresponding POSS plate kindly provided by Mike Irwin (RGO).

**Figure 6.** X-ray surface brightness contours in the 0.52 to 2.0 keV band overlaid onto the optical (POSS E) image. The size of the field is $50' \times 50'$; a redshift of 0.04 has been assumed in the conversion from angular to metric scale (indicated in the lower left corner). The X-ray image has been smoothed adaptively prior to contouring; the minimum number of counts over which smoothing occurred is 50. Contour levels are equally spaced on a logarithmic scale, starting at $6 \times 10^{-4}$ cts s$^{-1}$arcmin$^{-2}$ and increasing by a factor of two between adjacent contours. The contour labeling is in units of $10^{-3}$ cts s$^{-1}$arcmin$^{-2}$. The brightness of the background is $2.3 \times 10^{-4}$ cts s$^{-1}$arcmin$^{-2}$ so that the lowest contour level is a factor of 2.6 above the background; its S/N is more than $7\sigma$.

**Figure 7.** Hard-band X-ray surface brightness contours in the 0.52 to 2.0 keV band around HCG 94 (top) and A 2572 (bottom) overlaid onto the optical (POSS E) images. The size of the fields is $15' \times 15'$; a redshift of 0.04 has been assumed in the conversion from angular to metric scale (indicated in the lower left corner of each image). The X-ray images have been smoothed adaptively prior to contouring; the minimum number of counts over which smoothing occurred is 50. Contour levels are as in Fig. 6. The elongated feature south of A 2572 at $23^h 18^m 35^s$, $18°39'00''$ is a plate flaw.

emission from the northern subcluster in A 2572 (cf. lower right panel in Fig. 3) has a stellar counterpart in the optical.

• The point source west of the northern X-ray peak of A 2572 is prominent in both the hard and the soft X-ray image and most probably due to coronal emission from HD 219726, a late-type (G0) star of $m_v = 8.8$ at R.A. $23^h 18^m 4.45^s$, Dec. $+18°43'42.3''$ [Fig. 7 (bottom)].

• A second point source north-east of HCG 94 coincides with an unidentified faint object, possibly a high redshift QSO. The same holds for the brighter of the two sources at about $23^h 17^m 50^s$, $+18°51'$, i.e. some 10 arcmin north of A 2572 and HCG 94 featuring prominently in the foreground of Fig. 2. Here the most likely optical counterpart is bright enough to be classified as stellar.

• The X-ray emission from HCG 94 is centred on the two bright elliptical galaxies NGC 7578A and B [HCG 94a and b in Hickson (1993); galaxies f and g in Fig. 5].

• X-ray emission from two bright galaxies, NGC 7588 (galaxy 'p' just north-west of HD 219726) and NGC 7572 (galaxy 'b' 15 arcmin south-west of HCG 94) is detected on top of the diffuse ICM emission; both galaxies are cluster members and have $z = 0.0369$ and $z = 0.0446$, respectively.

• Additionally, at least one more of the spectrally hard sources around the main body of emission from HCG 94 can be associated with a group of galaxies whose radial velocities are, however, unknown, namely the patch of emission just west of NGC 7572 which coincides with a group of three or more galaxies at about $23^h 16^m 25^s$, $+18°28'$.

There are also a number of non-detections and unidentified X-ray sources that are worth mentioning.

• No X-ray emission from the foreground spiral UGC 12471 (galaxy 'a') is detected in our more than 20 ks exposure.

• With the exception of HD 219726 (see above), none of the four bright stars projected onto A 2572 are particularly X-ray bright.

• The two sources south of HD 219726 (seen best in Figs 1 and 2) do not coincide with any prominent object in the optical (note the clear separation from HD 219713 in the bottom panel of Fig. 7). Neither source shows discernible extent in its emission, but both are spectrally hard and not detected in the PSPC's soft energy band.

• The radio galaxy 'r', situated just inside the inner ring of the PSPC support structure, is not detected. However, it is located in a region where the diffuse cluster emission is still more than twice as strong as the signal expected from the galaxy from the correlation between X-ray flux and core radio flux (Edge & Röttgering 1995).

**Figure 8.** The regions of interest for the spectral X-ray analysis within the field of Fig. 1. See Fig. 6 for details about the underlaid contour map.

## 4 SPECTRAL X-RAY ANALYSIS

We analysed the spectral properties of the extended emission around HCG 94 and A 2572 in ten areas of special interest. They are

$A_0$  The core of HCG 94, a circular region of 1 arcmin radius centred on $23^h 17^m 12.0^s$, $+18°42'13''$.

$A_1$  An inner annulus from 1 to 3 arcmin around HCG 94

$A_2$  The eastern part (facing A 2572) of a second annulus from 3 to 10 arcmin around HCG 94; the opening angle is $120°$.

$A_3$  The western remainder (away from A 2572) of the same annulus.

$A_4$  The western part (opening angle $260°$) of an annulus from 10 to 17.5 arcmin around the centre of HCG 94.

$B_0$  The core of the northern part of A 2572, a circular region of 1 arcmin radius centred on $23^h 18^m 21.4^s$, $+18°43'54''$.

$B_1$  An annulus from 1 to 8 arcmin around the northern X-ray peak in A 2572, truncated at the dividing line to the southern peak and clipped at an off-axis angle of $19.5°$ to exclude emission affected by the PSPC window support structure.

$C_0$  The core of the southern part of A 2572, a circular region of 1 arcmin radius centred on $23^h 18^m 29.3^s$, $+18°41'31''$.

$C_1$  An annulus from 1 to 10 arcmin around the southern X-ray peak in A 2572, truncated at the dividing line to the northern peak and clipped at an off-axis angle of $19.5°$ to exclude emission affected by the PSPC window support structure.

$D$  The region in an annulus from 30 to 45 arcmin around the centre of the pointing which serves to determine the background for A and B.

Figure 8 shows these regions of interest (ROI) selected for a detailed spectral analysis overlaid on a contour plot of the smoothed X-ray emission in the $0.5 - 2$ keV band. Also shown are 11 compact sources which were blanked out using circular masks of 30 to 60 arcsec radius depending on the apparent size of the sources. The dotted line in Fig. 8 marks the region where obscuration by the detector's foil support structure causes the effective exposure time in the PSPC broad band to fall below 95 per cent of the unobscured value. Details of the photon statistics for the regions of interest are given in Table 2. The background for both the spectral and the spatial X-ray analysis (see Section 5) is taken from an annulus from 30 to 45 arcmin around the nominal pointing position. Figure 9 shows the selected area; note that the spokes of the PSPC window support structure and all apparent sources in this region have been masked out.

A quantitative description of the spectral differences between the emission from the regions shown in Fig. 8 is obtained by detailed spectral modelling. Prior to fitting any models, we group the data such that each energy bin contains at least 20 photons. All fits are performed in the energy range from 0.2 to 2.0 keV.

**Figure 9.** The region (shown at enhanced contrast) used for the determination of the background for both the spatial and the spectral analysis. The underlaid image shows the raw counts in the PSPC broad energy band from 0.1 to 2.4 keV.

We start with a four component model to parametrize the spectrum of the X-ray background. Scattered Solar X-rays are modelled as a diffuse thermal plasma at $kT = 0.124$ keV plus a single Oxygen K$\alpha$ line at 0.54 keV (Snowden & Freyberg 1993); on top of that comes the particle background which is assumed to follow a powerlaw with photon index 1.97 (Snowden et al. 1992). Finally, an extragalactic component from unresolved AGN is added in the form of a powerlaw absorbed by the Galactic column of neutral Hydrogen. (Allowing the absorption to float does not improve the overall fit and results in a value consistent with the Galactic one.) Fitting this model to the spectral data from the background region in the $0.2 - 2.0$ keV range, we find the best-fit amplitude of the Oxygen K$\alpha$ line to be consistent with zero. Dropping it from our model we obtain a fit of essentially unchanged quality with a $\chi^2$ value of 206.1, corresponding to a reduced $\chi^2$ value of $\chi^2/\nu = 1.23$ ($\nu$ is the number of degrees of freedom). The best fit value for the photon index of the extragalactic power law component is $2.16^{+0.09}_{-0.09}$ in good agreement with the value of 2.12 determined by Hasinger (1992) from deep ROSAT PSPC pointings.

The data from the nine regions of interest labeled $A_i$, $B_j$, $C_j$ ($i = 0, 4; j = 0, 1$) are fitted assuming the above model for the background contribution and various models for the ICM emission. In these fits, we fix the relative amplitudes of the three background components at the best fit values obtained in the background region (having corrected for vignetting beforehand), and set the power law slope of the extragalactic contribution to 2.12. The overall background normalization is then scaled to account for the differences between the areas of the data and background regions.

We model the ICM emission as thermal emission from a single temperature diffuse plasma using an updated version of the code described by Raymond & Smith (1977). Gas temperature ($kT$), metal abundance ($Z$), equivalent Hydrogen column density ($N_\mathrm{H}$), and a normalization factor are the four fit parameters for this model. The Solar abundances used in the plasma code are as in Allen (1973) with the Fe abundance being 6.3 where the Hydrogen abundance by number is 12 on a logarithmic scale. Any excess absorption over the Galactic value acts on the source spectrum and the extragalactic component of the background model alike. This model provides a satisfactory fit to the data with reduced $\chi^2$ values of about 1 in all regions. Table 3 summarizes the results of the spectral fits.

### 4.1 HCG 94

In the regions $A_i$ we find the metal abundance to be poorly constrained in general but consistent with $Z \sim 0.3\,Z_\odot$. However, there is a trend for both metal abundances and column densities to rise, and for gas temperatures to fall towards the core of HCG 94 (see Fig. 10). The vertical error bars in Fig. 10 correspond to a confidence level of 90 per cent; in the horizontal direction they represent the radial bin width. In the outermost region, $A_4$, which is dominated by background emission (cf. Table 2) we find the photon statistics to be too poor to yield meaningful results for all three fit parameters. The best-fit values and errors for this region were therefore obtained with the metal abundance frozen at $Z = 0.12\,Z_\odot$, the best fit value for the adjacent annulus $A_2 \cup A_3$.

| ROI | area (arcmin$^2$) | # of photons | # of photons above background | signal/ background |
|---|---|---|---|---|
| $A_0$ | 3.1 | 2453 | 2424 | 84.8 |
| $A_1$ | 25.1 | 5982 | 5751 | 24.8 |
| $A_2$ | 93.5 | 3978 | 3116 | 3.61 |
| $A_3$ | 185.3 | 8294 | 6586 | 3.85 |
| $A_4$ | 465.0 | 7560 | 3273 | 0.76 |
| $B_0$ | 2.5 | 1293 | 1260 | 37.7 |
| $B_1$ | 99.1 | 4660 | 3746 | 4.10 |
| $C_0$ | 3.1 | 1367 | 1338 | 46.8 |
| $C_1$ | 62.6 | 2784 | 2207 | 3.82 |
| D | 2787.8 | 25704 | n.a. | n.a. |

**Table 2.** Areas of and photon counts in the regions of interest shown in Fig. 8. The given numbers of photons and signal-to-background ratios are for the energy range from 0.2 to 2.0 keV used in the spectral fits.

**Figure 10.** Radial variations in the best fit column density (top) and the plasma temperature (bottom) from the spectral fits in the regions $A_0$, $A_1$, $A_2 \cup A_3$, and $A_4$. The dashed horizontal line represents the Galactic Hydrogen column density; its 90 per cent error range is marked by the dotted lines.

According to Fig. 10 the temperature of the ICM drops from about 3 keV at radii greater than 300 kpc to less than 2 keV in the inner 70 kpc, whereas, over the same radial range, the Hydrogen column density rises from about Galactic to 30 per cent excess ($\Delta N_H \sim 1 \times 10^{20}$ cm$^{-2}$). The significance of the radial gradients in both gas temperature and absorption can be assessed more clearly from Fig. 11 which shows the 68 and 90 per cent confidence contours for the parameters $kT$, $Z$, and $N_H$. Note that in $Z - N_H$ space the 90 per cent confidence contours for the three annuli hardly overlap.

In Fig. 11 we see an anti-correlation between metal abundance and absorption. This is due to the combination of a low-temperature plasma with the limited spectral resolution of the PSPC. A steepening in the spectral slope below 1 keV can be achieved equally well by raising either the column density, and thus the absorption ($\propto \exp[-n_\mathrm{H}\sigma(E)]$), or by increasing the metallicity which enhances line emission (around 1 keV) in a low temperature plasma.

In the interface region $A_2$ we find the temperature to be rather lower than the one of the plasma in the opposite region $A_3$. However, the uncertainties in all three fit parameters are large for these regions, so that the differences between $A_2$ and $A_3$ are only marginally significant (Fig. 12).

As far as its spectral X-ray properties are concerned, HCG 94 can thus be described as a fairly typical galaxy cluster (not group) of average luminosity, whose relaxed state is apparently not yet affected by the imminent merger with A 2572. The decrease of the ICM temperature and increase of absorption toward the core of HCG 94 is similar to what Allen & Fabian (1994) find in their X-ray study of the Centaurus cluster of galaxies which features about the same mass as HCG 94. Radial gradients in kT and $N_H$ as observed here also appear to be common in more massive clusters (Allen et al. 1995).

**Figure 11.** Contour plot (68 and 90 per cent confidence) of the allowed regions in $kT - N_H$ (top) and $Z - N_H$ space (bottom) from the spectral fits in the regions $A_0$, $A_1$, and $A_2 \cup A_3$. The best fit values for each region are marked by the respective ROI label. The dashed horizontal line represents the Galactic Hydrogen column density; the corresponding 90 per cent error limits are shown as dotted lines.

**Figure 12.** Contour plot (68 and 90 per cent confidence) of the allowed regions in $kT - N_H$ (top) and $Z - N_H$ space (bottom) from the spectral fits in the regions $A_2$ and $A_3$. The best fit values for each region are marked by the respective ROI label. The dashed horizontal line represents the Galactic Hydrogen column density; the corresponding 90 per cent error limits are shown as dotted lines.

**Figure 13.** Contour plot (68 and 90 per cent confidence) of the allowed regions in $kT - N_H$ space from the spectral fits in the regions B and C. The best fit values for each region are marked by the respective ROI label. The dashed horizontal line represents the Galactic Hydrogen column density; the corresponding 90 per cent error limits are shown as dotted lines.

### 4.2 A 2572

In the regions B (A 2572 north) and C (A 2572 south) the spectral analysis is considerably more complicated due to the poorer photon statistics and the overlap between the northern and southern source. We assume a redshift of $z = 0.155$ for the northern cluster (cf. Section 6).

Fitting a single temperature plasma to the spectrum obtained in region B we find its temperature of about 8 keV to be much higher than the one in region C ($\sim 2.5$ keV), which supports the hypothesis that the emission originates from a more distant cluster of much greater luminosity than A 2572. Even for a temperature as low as about 5 keV (which, according to Fig. 13, is a lower limit at the 90 per cent confidence level) the northern cluster would have to be at a redshift of greater than 0.08 in order to have its luminosity fall within $3\sigma$ of the $L_X - kT$ relation for clusters. Since the metallicity could not be constrained in the spectral fit for the northern cluster, it was frozen at 30 per cent of the Solar value. Including a further plasma component to account for spectral contamination from the respective other cluster did not yield a significant amplitude for such a second component in either region. The lack of overlap between the $\chi^2$ contours in Figure 13 shows these fit results to be clearly inconsistent with the assumption of a common relaxed gaseous envelope around both clusters.

Figure 14 and Table 3 summarize the results of the spectral fits in region B. Again, we find the temperature in the core of the emission to be lower than in the outer regions. The radial temperature variation seen in Fig. 14 is, however, rather one of lower limits, as, due to the softness of the ROSAT energy range, the high temperatures derived in region B are all poorly constrained at the top end.

In region C the best-fit ICM temperature of 2.4 keV is obtained for a metal abundance in excess of the Solar value. Although a high metallicity is favoured also in the outer region $C_1$, the high overall value is primarily due to the peculiarities of the spectrum in the immediate surrounding ($C_0$) of the central cluster galaxy, NGC 7597. A single temperature plasma model with three free parameters yields an acceptable fit to the data in region $C_0$ with $kT = 1.6^{+0.6}_{-0.2}$ keV and $Z = 2.2^{+2.8}_{-1.1} Z_\odot$ ($\chi^2_\nu = 0.72$); however, excess emission at energies below 0.5 keV causes the best-fit column density to come out unphysically low at $3.2^{+0.9}_{-1.2} \times 10^{20}$ cm$^{-2}$, which is inconsistent with the Galactic value at the greater 99 per cent confidence level. (In the outer region $C_1$ the best-fit column density is consistent with the Galactic value.) Freezing the absorption at the Galactic value and fitting only the 0.5 to 2 keV range affects the best-fit temperature only marginally ($kT = 1.6^{+0.3}_{-0.2}$ keV), but causes the metallicity to drop to $Z = 1.6^{+2.0}_{-0.8} Z_\odot$ while the quality of the fit remains essentially unchanged ($\chi^2_\nu = 0.75$).

**Figure 14.** Contour plot (68 and 90 per cent confidence) of the allowed regions in $kT - N_H$ space from the spectral fits in the regions $B_0$ and $B_1$. The best fit values for each region are marked by the respective ROI label. The dashed horizontal line represents the Galactic Hydrogen column density; the corresponding 90 per cent error limits are shown as dotted lines.

**Figure 15.** The spectrum (convolved with the instrument response) obtained around NGC 7597 (region $C_0$) and the best fitting single temperature plasma model. A two parameter least-squares fit with Galactic absorption is performed over the 0.5 to 2.0 keV range. The bottom panel shows the residuals. The extrapolation down to 0.2 keV shows the excess emission at low energies.

Figure 15 shows the spectrum in region $C_0$ and the best fitting single temperature thermal plasma model. Note that although the overall fit is acceptable, the single temperature RS model has difficulties reproducing the observed spectrum at energies around below 0.4 keV and possibly also around 0.8 keV.

The ROI $C_0$ includes the very central region around the southern X-ray peak in A 2572 where a significant decrease in surface brightness was found in the image analysis (cf. Fig. 3 right). However, this feature which might be due to intrinsic absorption over the central 30 kpc is not in direct conflict with the observation of excess soft emission in our spectral analysis where a region of more than 130 kpc diameter is investigated.

The best-fitting ICM temperature of 4.3 keV quoted in Table 3 for the outer region $C_1$ should not be taken at face value, as the whole region is likely to be considerably contaminated by spectrally hard emission from the plasma in region $B_1$. However, with the data at hand we are not able to disentangle the two components even if we freeze the column density at the Galactic value.

In summary of the spectral analysis of region C we find it to be significantly cooler than the northern cluster, but are unable to further constrain the parameters governing the ICM properties due to the ubiquitous contamination from the neighbouring hot northern cluster, especially at greater radii.

### 4.3 The $\beta$ parameter

Given the uncertainties in membership of individual galaxies to the various clusters we expect the velocity dispersions quoted in Section 3 to be somewhat insecure. In order to test our prediction that the true values should in fact be lower than the measured ones, we compare spectral and imaging determinations of the $\beta$ parameter. The beta parameter is defined as the ratio of the specific kinetic energies of the galaxies to that of the intracluster gas:

$$\beta = \frac{\mu m_p \sigma^2}{kT}$$

where $\mu$ is the mean molecular weight of the ICM gas, $m_p$ is the proton mass, $\sigma$ is the cluster galaxies' velocity dispersion, and $T$ is the ICM gas temperature.

Setting $\mu$ to 0.6 and using the $\sigma$ values of 663 and 480 km s$^{-1}$ for HCG 94 and A 2572 (south) derived in Section 3, we find the above spectral ICM temperatures to yield beta values of 0.98 and 0.72 for HCG 94 and A 2572, respectively. (A temperature of 2 keV has been assumed for A 2572.) If both values for the radial velocity dispersion are actually upper limits, so are these beta values. The $\sigma$ value of 479 km s$^{-1}$ obtained by Hickson for just the seven galaxies of HCG 94 proper provides a lower limit of $\beta = 0.51$ for the cluster HCG 94. In the following section these figures will be compared to the respective values from the imaging analysis.

| ROI | $\chi^2$ | $\chi^2/\nu$ | $kT$ keV | $N_H$ $10^{20}$ cm$^{-2}$ | $Z$ $Z_\odot$ |
|---|---|---|---|---|---|
| $A_0$ | 89.1 | 1.00 | 1.7 (1.4,2.2) | 6.6 (5.7,7.6) | 0.56 (0.31,1.03) |
| $A_1$ | 143.4 | 1.00 | 2.4 (1.9,3.2) | 6.0 (5.4,6.6) | 0.30 (0.13,0.57) |
| $A_2$ | 134.2 | 1.10 | 2.4 (1.7,3.7) | 5.7 (4.9,6.7) | 0.11 (0.00,0.44) |
| $A_3$ | 138.5 | 0.87 | 3.4 (2.6,4.8) | 4.9 (4.5,5.4) | 0.10 (0.00,0.46) |
| $A_2 \cup A_3$ | 158.7 | 0.94 | 2.8 (2.3,3.6) | 5.1 (4.7,5.5) | 0.12 (0.00,0.32) |
| $A_4$ | 162.2 | 1.10 | 2.8 (1.7,5.9) | 5.7 (5.1,6.4) | 0.12 (*frozen*) |
| $B_0$ | 44.0 | 0.88 | 4.2 (2.6,8.8) | 4.4 (3.8,5.0) | 0.30 (*frozen*) |
| $B_1$ | 168.5 | 1.23 | 15.1 (6.6,uc) | 4.2 (3.7,4.7) | 0.30 (*frozen*) |
| $B_0 \cup B_1$ | 159.5 | 1.07 | 8.5 (5.3,20.8) | 4.2 (3.9,4.6) | 0.30 (*frozen*) |
| $C_0$ | 32.3 | 0.75 | 1.6 (1.4,1.9) | 4.7 (*frozen*) | 1.59 (0.8,3.6) |
| $C_1$ | 90.0 | 0.96 | 4.3 (2.3,8.8) | 4.8 (3.9,5.8) | 1.21 (0.2,uc) |
| $C_0 \cup C_1$ | 128.4 | 1.03 | 2.4 (1.8,4.1) | 4.3 (3.7,5.0) | 1.56 (0.81,uc) |

**Table 3.** Best fit values and 90 per cent confidence ranges for the parameters of a thermal plasma model fitted to the spectra from the regions of interest. The fitted energy range is 0.2 to 2.0 keV for all regions except $C_0$ where only the range from 0.5 to 2.0 keV was taken into account. 'uc' stands for 'unconstrained'.

**Figure 16.** The regions of interest for the spatial analysis within the field of Fig. 1. See Fig. 6 for details about the underlaid contour map.

**Figure 17.** The radial surface brightness distribution in the 0.5 to 2 keV band around HCG 94. The solid line shows a King model fitted to the data in the $r > 150$ kpc range to exclude the cooling flow emission. The background (shown as a dotted line) has been taken into account in the fitting.

## 5 SPATIAL X-RAY ANALYSIS

For a more quantitative analysis of the X-ray surface brightness distributions depicted in Figs. 1 to 4 we extract the photons detected within three regions around the main peaks of the ICM emission.

1 A circular region of 17.5 arcmin radius centred on the peak of the emission from HCG 94; in a cone of 100 degrees opening angle towards A 2572 only data within 10 arcmin radius are considered to avoid the interface region with A 2572.
2 A semi-circular region of 8 arcmin radius centred on the northern peak of the emission from A 2572.
3 A semi-circular region of 8 arcmin radius centred on the southern peak of the emission from A 2572.

Figure 16 shows these regions overlaid on the same contours as in Fig. 8. The background in the $0.5 - 2$ keV band is determined from the same region D used as a background field in the spectral analysis (cf. Fig. 9).

In the three regions of interest we parametrize the projected spatial ICM distribution by modified King models (King 1962) of the type

$$\sigma(r) = \sigma_0 \left[1 + (r/r_c)^2\right]^{-3\beta+1/2} + \sigma_{\text{bkg}}$$

which are fitted to the radial surface brightness profiles in the outer regions of the three sources where we can expect the effects of excess absorption and/or cooling near the cluster cores to be negligible. While this procedure ensures that the outer slope of the profile, i.e. the beta parameter, can be determined with reasonable accuracy, it entails systematic uncertainties for the best-fit core radius. These are hard to quantify but can be assessed by changing the radial range over which the fit is performed; in any case they outweigh by far the statistical errors. In the following, we quote only the best-fit values but explore the allowed range for the value of the more robust parameter, $\beta$ (cf. Section 4.3), by altering the fit range.

We then run a deprojection algorithm on the data from the PSPC's hard energy band that allows the true three-dimensional geometry of the cluster gas properties to be reconstructed from the projected surface brightness distribution. The deprojection technique was developed for the analysis of imaging X-ray data of clusters of galaxies by Fabian et al. (1981) and has since then been used commonly in the analysis of data taken with the EINSTEIN, EXOSAT, and ROSAT observatories. A description of the deprojection method can be found in White's (1995) catalogue of deprojection results from EINSTEIN observations of 188 clusters of galaxies.

### 5.1 HCG 94

Figure 17 shows the radial surface brightness distribution obtained in region 1, the area around HCG 94. The overlaid King model is the result of a least-squares fit to the data in which the inner 150 kpc, i.e. the cooling flow region (see Section 4), have been ignored. The best fit ($\chi^2_\nu = 1.3$) is obtained for a core radius $r_c = 170$ kpc and a slope of $\beta = 0.58$. However, as can be seen from Fig. 17, the profile steepens somewhat with radius, and best-fit beta values of up to 0.92 can be obtained when the radius of the core region excluded from the fit is gradually increased to 0.5 Mpc.

Figure 18 summarizes the results of the deprojection analysis for region 1, i.e. the area around HCG 94. We again use the King approximation to the gas density distribution in an isothermal sphere to model the observed surface brightness profile, and require the gas temperature in the outer regions of the cluster to be consistent with the value of $kT = 2.8$ keV found in the spectral analysis (see Section 4). The latter criterion can be met only for a cluster velocity dispersion of $\sigma \sim 500$ km s$^{-1}$. As far as the cluster core radius is concerned, values between 100 and 250 kpc can all be reconciled with a gas temperature of about 2.8 keV at $r \sim 1$ Mpc; however, the more extreme values in this range are increasingly in conflict with the projected temperature profile determined from the spectra (see Section 4; note that Fig. 18 shows the *deprojected* temperature profile). The results in Fig. 18 are for a core radius of 170 kpc and $\sigma = 500$ km s$^{-1}$. The total mass of HCG 94 within 1 Mpc from the cluster core amounts to more than $1.3 \times 10^{14}$ M$_\odot$; its X-ray luminosity in the 0.1 to 2.4 keV band is $1.3 \times 10^{44}$ erg s$^{-1}$. These values are in good agreement with the general $L_X - T$ and $\sigma - T$ relations for clusters of galaxies (David et al. 1993, White 1995).

We find HCG 94 to host a moderate cooling flow featuring a mass deposition rate of $\dot{M} = 45^{+12}_{-27}$ M$_\odot$ yr$^{-1}$ (errors correspond to the 10th and 90th percentile) within a cooling radius of $120^{+30}_{-70}$ kpc. The mass deposited by the cooling flow in a Hubble time is roughly consistent with the small but significant excess column density (see Section 4.1) found in the spectral analysis [see also White et al. (1991)]. The cumulative baryon fraction reaches a minimum of 10 per cent at a radius of 150 kpc but increases steadily with radius to a value of more than 28 per cent at $r = 1.15$ Mpc. This figure is even higher than the value of 20 per cent found at the same radius for the Coma cluster (White & Frenk 1991) and underlines the problem of baryon overdensities encountered in many X-ray observations of clusters of galaxies (White & Fabian 1995).

### 5.2 A 2572

For the regions 2 and 3 of Fig. 16 the spatial analysis is complicated by the fact that the emission profiles of the two sources overlap considerably. Also, the extended emission from HCG 94 has to be taken into account, so that the simple assumption of an undisturbed, spherically symmetric emission profile (as it was made for HCG 94) is no longer justified. In order to determine at least approximate radial profiles for the two clusters next to the nominal position of A 2572, we subtract the emission from HCG 94 from the surface brightness distribution in the regions 2 and 3. Having done so, we attempt to disentangle the contributions from the southern and northern clusters by fitting King models to the radial profiles in the two regions and subtracting the contribution from either one source in the region of the other. Repeating this procedure in an iterative loop quickly leads to convergence in the parameter values for the two King profiles. However, we can not expect the profiles thus derived to be as accurate as the one of HCG 94 as any inhomogeneities in the surface brightness distributions and any deviation from spherical symmetry in either of the two sources will lead to systematic errors in the corrected profiles. These will be most apparent in the outermost regions where the surface brightness values of all components are low and, in places, of similar amplitude.

Figure 21 shows the surface brightness profiles thus obtained for the two X-ray peaks in A 2572. The best-fit parameters are $r_c = 178$ kpc, $\beta = 0.59$, and $r_c = 30$ kpc, $\beta = 0.45$ for the northern and the southern source, respectively (redshifts $z$ of 0.155 and 0.03893 are assumed). As for HCG 94, the radial fit for the northern source has been performed for $r > 150$ kpc; however, for A 2572 (north) the best-fitting beta value does not vary greatly with radius. Only for cutoff radii greater than 0.5 Mpc does the profile flatten slightly, and $\beta$ approaches 0.54. As for HCG 94, we hence find the imaging $\beta$ value of the northern cluster to be close to the canonical value of 0.64 found in imaging observations of clusters of galaxies with the EINSTEIN Observatory (Jones & Forman 1984). For the southern source, i.e. A 2572 proper, we exclude only the central 32 kpc, which is where we find the emission in the PSPC hard band to be absorbed in Fig. 3. The same absorption feature is clearly visible in the radial surface brightness distribution. The best-fitting beta value is not sensitive to variations in the lower end of the fit range from 32 to 150 kpc; we always find values between 0.43 and 0.45 (the final three bins are excluded in these fits). These are significantly lower, i.e. the surface brightness profile is significantly shallower, than that of rich clusters of galaxies and more like that of galaxy groups (e.g. Ponman & Bertram 1993, Böhringer 1994, Pildis et al. 1995).

Figure 19 summarizes the deprojection results obtained in region 2 when a redshift of 0.155 is assumed. The cluster velocity dispersion and core radius are fixed at $\sigma = 800$ km s$^{-1}$ and $r_c = 200$ kpc, respectively; the temperature profile is adjusted such that it qualitatively reproduces the values determined in the spectral fits detailed in Section 4. Although at radii greater than 1 Mpc the uncertainties in the subtraction of the surface brightness contributions from both HCG 94 and A 2572 (south) are reflected in the run of almost all parameters in Fig. 19, the main deprojection results are robust. Again the values for luminosity ($3.7 \times 10^{44}$ erg s$^{-1}$ in the 0.1 to 2.4 keV band), velocity dispersion, and gas temperature are consistent with the correlations found between these quantities in studies of larger cluster samples. Although ill-constrained, the mass deposition by the central cooling flow is significant; we find $\dot{M} = 43^{+41}_{-16}$ M$_\odot$ within $r_{\rm cool} = 80^{+80}_{-30}$ kpc. The gravitational mass within 1 Mpc exceeds $3 \times 10^{14}$ M$_\odot$. As in HCG 94, the baryon fraction rises with radius and reaches 20 per cent at $r = 1.5$ Mpc.

The deprojection results for region 3 (the southern cluster) are shown in Fig. 20. An isothermal model with a core radius of 30 kpc has been assumed for the gravitational potential; the radial velocity dispersion is set to 350 km s$^{-1}$. As before, the caveat applies that systematic effects introduced by imperfections in the subtraction of emission from the northern cluster or in the flat fielding of the raw counts in the immediate neighbourhood of the PSPC support structure may affect our results. Again we find the baryon fraction to increase with radius; at $r = 500$ kpc a value of almost 15 per cent is attained. Within a radius of about 100 kpc around NGC 7597 the ICM's cooling time falls below a Hubble time resulting in a cooling flow onto the central galaxy of $\dot{M} = 14^{+4}_{-9}$ M$_\odot$ yr$^{-1}$ within $r_{\rm cool} = 90^{+24}_{-40}$ kpc. The gravitational mass within 500 kpc is found to be $3.5 \times 10^{13}$ M$_\odot$, the luminosity in the 0.1 to 2.4 keV band is $3.2 \times 10^{43}$ erg s$^{-1}$. When compared to poor clusters of similar mass, the gas temperature of 2 keV used in the analysis is somewhat high but not yet in conflict with other observations.

### 5.3 $\beta$ parameter values

The imaging $\beta$ values derived above are consistent with the constraints obtained from the spectral analysis (see Section 4) where we found $0.51 < \beta_{\rm HCG\,94} < 0.98$ and $\beta_{\rm A\,2572} < 0.72$. Due to the uncertainties (both statistical and systematical) in the determination of the cluster velocity dispersions, the ICM gas temperatures, and the slope of the King profile fitted to the projected X-ray surface brightness distributions (the latter two parameters are particularly difficult to determine for the two-component emission around A 2572), no more than qualitative agreement can be expected anyway. However, the comparison of the results from the spectral and the imaging analysis can be used to test the self-consistency of optical and X-ray parameters.

Adopting the beta values from the imaging analysis we can actually infer velocity dispersions for all three clusters. We find 510 km s$^{-1}$ for HCG 94, in good agreement with the value of about 500 km s$^{-1}$ required by the deprojection procedure. However, this value is considerably lower than the velocity dispersion of 663 km s$^{-1}$ derived for HCG 94 in Section 3 which supports our claim that our simplistic interpretation of the measured radial velocities yields values that are most probably overestimated.

**Figure 18.** Results of the deprojection analysis of the surface brightness distribution of HCG 94 in the PSPC's 0.5 to 2.0 keV band. All distributions are shown as a function of radial distance from the centre of the X-ray emission; a redshift of $z = 0.04218$ has been assumed in the scaling from angular to metric distances. From left to right and top to bottom: X-ray surface brightness in the 0.5 to 2.0 keV range; gas pressure; cumulative bolometric X-ray luminosity; ICM gas temperature; electron density; cooling time; cumulative gas and gravitational mass; cumulative mass deposition rate. The dotted lines in the fourth and sixth panel mark the required ICM gas temperature of 2.8 keV determined in the spectral analysis, and the age of the Universe, respectively. Errors are $1\sigma$ except for k$T$, $T_{\rm cool}$, and $\dot M$ where the 10$^{\rm th}$ and 90$^{\rm th}$ percentiles are shown.

**Figure 19.** Results of the deprojection analysis of the surface brightness distribution of the northern X-ray peak of A 2572 in the PSPC's 0.5 to 2.0 keV band. All distributions are shown as a function of radial distance from the centre of the X-ray emission; a redshift of $z = 0.155$ has been assumed in the scaling from angular to metric distances. The individual panels are as in Fig. 18. The dotted lines in the fourth and sixth panel mark the required ICM gas temperature of approximately 9 keV determined in the spectral analysis, and the age of the Universe, respectively.

**Figure 21.** The radial surface brightness distribution in the 0.5 to 2 keV band around the northern (top) and the southern source around A 2572. The solid line shows a King model fitted to the data in the $r > 150$ kpc (northern source) and the $r > 32$ kpc range (southern source). The background (shown as a dotted line) has been taken into account in the fitting.

For A 2572 north, we find 860 km s$^{-1}$ if a gas temperature of 8 keV is assumed. Since no radial velocities are available that could unambiguously be related to this cluster we cannot compare this figure to an actual measurement. It would, however, be consistent with the derived luminosity and gas temperature.

The range of velocity dispersion derived from the imaging $\beta$ for A 2572 proper is $330 < \beta ({\rm km\ s}^{-1}) < 380$ if ICM gas temperatures between 1.5 and 2 keV are assumed. Again these values are lower than the formal measured value of 480 km s$^{-1}$ presented in Section 3 but consistent with what would be expected for a loose group or a very poor cluster with mass and gas temperature as determined above.

## 6 DISCUSSION

In the PSPC data we find the X-ray emission around the optical position of A 2572 to be dominated by three bright and clearly extended sources, one of which coincides with the nearby compact group of galaxies HCG 94. Using both the X-ray spatial and the spectral properties of the emission in the region around HCG 94, we show that HCG 94 is in fact the core of a proper galaxy cluster featuring a gravitational mass of more than $1 \times 10^{14}$ M$_\odot$ within 1 Mpc, and an unabsorbed X-ray luminosity of $1.5 \times 10^{44}$ erg s$^{-1}$ in the 0.1 to 2.4 keV band. Both the ICM gas temperature and the column density determined in spectral fits exhibit a significant radial gradient suggesting that HCG 94 is a cooling flow cluster. A spatial deprojection analysis yields a mass deposition rate of 45 M$_\odot$ yr$^{-1}$ for the core of HCG 94.

The other two bright and extended sources in the central part of the PSPC's field of view are only separated by about 3 arcmin and coincide with A 2572 itself. The southern of these is perfectly centred on the elliptical galaxy NGC 7597 which is a cluster member at $z = 0.0376$ and the dominant galaxy in A 2572.

From the comparison of the optical and X-ray images there can be no doubt as to the cluster nature of the northern source either. However, none of the galaxies apparently associated with the emission has a measured redshift, so that we can not be sure they are actually members of A 2572. Equally worrying is the lack of a dominant bright elliptical galaxy close to the X-ray peak; if the galaxies visible on the POSS plate were at $z \sim 0.04$, i.e. constituted a subcluster of A 2572, then one would expect its dominant galaxy to be similar in brightness to NGC 7597 or NGC 7602, which is clearly not the case. Also there are no signs of interaction between the two systems in the X-ray as would be expected for such apparent substructure; the surface brightness profiles appear almost totally undisturbed. An alternative interpretation of the data at hand would be that A 2572 is in fact a superposition of two clusters both in the optical and in the X-ray, with the northern cluster being much more distant than $z \sim 0.04$. This would explain the narrow angle tail (NAT) radio morphology of the background galaxy 'r' (cf. Table 1 and Fig. 5) if we assume that its redshift ($z = 0.1547$) is also the redshift of this northern, more distant cluster. This scenario would place galaxy 'r' at a distance of 1.2 Mpc from the (northern) cluster core which is well within the 0.5 Abell radii O'Dea & Owen (1985) chose for the selection of their sample of NATs in clusters of galaxies. The X-ray spectrum from this region strongly supports this hypothesis, as we find the ICM in the northern cluster to be hotter than that in the southern component (A 2572). The true difference in temperature can in fact be expected to be even greater, as the superposition of the two emission regions tends to reduce the observed (projected) temperature gradient. An ICM temperature of about 7–8 keV for the northern cluster (as opposed to about $\sim 2$ keV for the southern one) would be unreasonably high when compared to the cluster's 0.1 to 2.4 keV luminosity of $3 \times 10^{43}$ ($z \sim 0.04$ assumed). If, however, the cluster was indeed at $z = 0.155$ and, accordingly, featured an X-ray luminosity of $4 \times 10^{44}$ erg s$^{-1}$, such a high temperature would be perfectly consistent with the $L_X - $k$T$ relation for galaxy clusters (e.g., David et al. 1993).

On the other hand, such a superposition appears very unlikely purely on statistical grounds. The northern peak in A 2572 features an X-ray flux of about $4 \times 10^{-12}$ erg cm$^{-2}$ s$^{-1}$ (0.1–2.4 keV). From the log $N -$ log $S$ relation for Abell and ACO clusters of galaxies detected in the ROSAT All-Sky Survey (Ebeling 1993, Ebeling et al. 1995) we find the probability of serendipitously finding a cluster of this X-ray brightness within a circle of 5 arcmin radius around any given point to be less than $5 \times 10^{-4}$. Still, in view of the striking (if circumstantial) evidence for a superposition, we shall mean only the southern cluster when referring to A 2572 in the following.

Given the projected proximity of A 2572 and HCG 94, what is the evidence that the two systems are in the process of merging? The difference in the mean radial velocities of the two cluster components (A 2572 and HCG 94) of 935 km s$^{-1}$ is of the order of the clusters' velocity dispersions. However, the number of measured redshifts in the whole complex is still too small to disentangle the galaxy populations of the two clusters. Consequently, the observed difference in the clusters' mean radial velocities is probably underestimated while the individual velocity dispersions are likely to be overestimated. As far as the three-dimensional arrangement of the two components is concerned the truth lies probably somewhere between the three extreme cases discussed in the following.

**Figure 20.** Results of the deprojection analysis of the surface brightness distribution of the southern X-ray peak of A 2572 in the PSPC's 0.5 to 2.0 keV band. All distributions are shown as a function of radial distance from the centre of the X-ray emission; a redshift of $z = 0.039$ has been assumed in the scaling from angular to metric distances. The individual panels are as in Fig. 18. The dotted lines in the fourth and sixth panel mark the required ICM gas temperature of approximately 2 keV determined in the spectral analysis, and the age of the Universe, respectively.

- One extreme would be a fly-by scenario in which both systems are at the same distance from the observer and the true three-dimensional separation is in fact the projected one, i.e. 1.1 Mpc. The system would be marginally bound but the extreme angular momentum of this configuration makes it rather implausible.

- If attributed entirely to the Hubble flow, the difference in the cluster redshifts would formally correspond to a separation along the line of sight of 19 Mpc (as compared to 1.1 Mpc in projection). The escape velocity at this separation amounts to less than a third of the observed relative velocity. Hence, the system could not be gravitationally bound in this picture.

- Alternatively, the complex A 2572/HCG 94 could already have decoupled completely from the Hubble expansion. The relative velocity of the two clusters would then be the result of their mutual gravitational attraction, which implies that, despite its higher redshift, it is actually HCG 94 that is the more nearby system. In the simple picture where we assume the two clusters to be point masses falling towards each other from infinity with zero angular momentum, the measured relative velocity translates into a spatial separation along the line of sight of about 2 Mpc, i.e. the Abell radii of the two clusters would overlap not just in projection but in fact also in three dimensions.

Although neither of these scenarios is totally compelling, it appears safe to assume that we are witnessing the early stages of the merger between A 2572 and HCG 94 as the metric three-dimensional separation is probably still a few Mpc. If we adopt this hypothesis, should we be seeing signs of interaction between the two components? Although, in the PSPC data, the intracluster gas in HCG 94 can be traced out to beyond 1 Mpc before the X-ray emission in the 0.5 to 2 keV band becomes indistinguishable from the background, the deprojected electron density falls below $10^{-4}$ cm$^{-3}$ beyond 1 Mpc. Given that the sound speed in the ICM, $v_s \sim 350 \, (kT/\text{keV})^{1/2}$ km s$^{-1}$, amounts to some 600 km s$^{-1}$ for HCG 94 and probably less for A 2572, the collision of the gas masses at the merger front should occur at velocities that are only mildly supersonic. However, at such low gas densities even a much stronger shock could increase the density at the shock front only to some $10^{-4}$ cm$^{-3}$. The resultant changes in the X-ray surface brightness would be comparable to the background fluctuations in our pointing and thus not observable at the available sensitivity level. The lack of direct evidence of interaction between the two clusters is therefore not in conflict with the merger picture.

As far as HCG 94 proper is concerned, the misclassification as a group was possible only because of the unusual compactness of the cluster core in the optical and the overall spatial segregation of bright and faint galaxies. No more than nine galaxies with magnitudes between $m_1$ and $m_1 + 3$ are found within a radius of $\sim 700$ kpc around the cluster centre. Six of these have been confirmed as cluster members; five are the brightest galaxies in the original group HCG 94, i.e. concentrated in the innermost 150 kpc of the cluster. Although comparable to regular Abell-type clusters of galaxies in terms of X-ray luminosity and gravitational mass, HCG 94 as a cluster is thus very unusual in its optical properties. Detailed optical follow-up observations of this system are planned.

The discovery that this group is in fact the core of a cluster raises the important question of how many of the HCGs are truly isolated systems. Recent studies (Ramella et al. 1994 and Rood & Struble 1994) show that there are very few truly isolated compact groups, and that the large majority of HCGs are embedded in richer, loose groups. However, the galaxy density of these neighbouring environments is usually low.

Although HCG 94 is thus anything but typical for HCGs in general, there is evidence that some fraction of Hickson's Compact Groups may be, at least loosely, associated with proper clusters, i.e. much richer systems, in what could be called an orbital mode. The only nearby ($z \le 0.02$) example of such a configuration is HCG 48 which was found to be a satellite of A 1060 (Rood and Struble 1994, see also Ebeling et al. 1994 for an X-ray image). The statistical study by Ebeling and co-workers suggests that almost 15 per cent of all HCGs can be expected to be in gravitational contact with Abell/ACO clusters in their near neighbourhood.

## 7 SUMMARY

Appearances can be deceptive. Contrary to historical classification, Hickson's Compact Group # 94 is not a compact group but rather the core of a proper cluster of galaxies. Also, A 2572 is not a (poor) cluster of galaxies but rather a loose group of galaxies. What, in the X-ray, appears to be the double core of A 2572 and a striking example of subclustering turns out to be most likely a superposition of A 2572 proper and a background cluster which we tentatively place at a redshift of $z = 0.155$. All these optical misclassifications were resolved by means of a spectral-imaging analysis of the ROSAT PSPC observation of the system.

However, it is not only the impressive amount of optical confusion associated with A 2572/HCG 94 that makes this system worth studying. HCG 94 is a prototypical example of a low-mass cooling flow cluster exhibiting all the classical characteristics such as radial gas temperature, metallicity and absorption gradients. Although these trends are clearly present, their interpretation from the PSPC data is slightly ambiguous; the ASCA observation of HCG 94 carried out in December 1994 will allow these issues to be resolved.

The cluster HCG 94 is the most extreme misclassification in Hickson's catalogue detected so far. Our findings underline the necessity for deeper optical observations of optically selected galaxy groups in conjunction with X-ray studies. HCG 94 also features unusually low galaxy counts for its mass and X-ray luminosity and exhibits a striking spatial segregation of bright and faint galaxies. A combined spectro-photometric investigation of the whole complex A 2572/HCG 94 is planned in order to establish the optical luminosity function of the cluster galaxies and obtain accurate redshifts and radial velocity dispersions for all three clusters in the field of view of our pointing.

The discovery of an X-ray luminous and thus massive cluster that is inconspicuous in the optical (and has in fact been detected only thanks to its compact core) suggests that past and present studies of clusters based on optically selected systems are in danger of systematically undersampling the true cluster population. Representative cluster samples can, however, be expected to emerge from the ROSAT All-Sky Survey.

As far as the loose group A 2572 is concerned, a more detailed and accurate investigation into its properties, in particular the run of metallicity and absorption near NGC 7597, is currently hampered by the ubiquitous contamination from the background cluster just north of it. Again, the ASCA observation of A 2572 should help to clarify the situation.

## 8 ACKNOWLEDGEMENTS

We would like to thank Mike Irwin for re-scanning the POSS E plate with the Automatic Plate Measuring facility at IoA, and Michael Ledlow who has been instrumental in resolving the confusion about the true redshift of the NAT radio galaxy south-east of A 2572. The following non-commercial software has been used in the data analysis: IDL ASTROLIB, FTOOLS, XSPEC – thanks to all contributors and maintainers of these packages. Thanks go also to S.W. Allen, W.N. Brandt, A.C. Edge, and C.S. Reynolds, as well as the referee, T.J. Ponman, discussions with whom have helped to improve this paper. H.E. acknowledges support by a European Union HCM Fellowship; D.A.W. acknowledges support from PPARC.

**REFERENCES**

Abell G.O., 1958, ApJS, 211, 3
Abell G.O., Corwin H.G., Olowin R.P., 1989, ApJS, 70, 1
Allen C.W., 1973, *Astrophysical Quantities*, The Athlone Press, London
Allen S.W. & Fabian A.C., 1994, MNRAS, 269, 409
Allen S.W., Fabian A.C., Edge A.C., Böhringer H., White D.A., 1995, MNRAS, in press
Beers T.C., Forman W., Huchra J.P., Jones C., Gebhardt K., 1991, AJ, 102, 1581
Böhringer H., in *Cosmological Aspects of X-Ray Clusters of Galaxies*, W.C. Seitter (ed), Kluwer Academic Publishers, 1994, p. 123
Briel U.G., Henry J.P., Schwarz R.A., Böhringer H., Ebeling H., Edge A.C., Hartner G.D., Schindler S., Trümper J., Voges W. 1991, A&A, 246, L10
Chincarini G & Rood H.J., 1972, AJ, 77, 4
David L.P., Slyz A., Jones C., Forman W., Vrtilek S.D., Arnaud K.A., 1993, ApJ, 412, 479
de Vaucoulers G., de Vaucouleurs A., Corwin H.G, et al., Third Reference Catalogue of Bright Galaxies, Springer 1991
Dressler A. & Shectman S.A., 1988, AJ, 95, 985
Ebeling H., 1993, PhD thesis, MPE report 250
Ebeling H., Voges W., Böhringer H., 1994, ApJ, 436, 44
Ebeling H., Edge A.C., Böhringer H., Voges W., Huchra J.P., Fabian A.C., Briel U.G., Allen S.W., 1995, in preparation
Edge A.C. & Röttgering H., 1995, submitted to MNRAS
Fabian A.C., Hu E.M., Cowie L.L., Grindlay J., 1981, ApJ, 248, 47
Forman W., Bechtold J., Blair W., Giacconi R., Van Speybroeck L., Jones C., 1981, ApJ, 243, L133
Forman W. & Jones C., in *Cosmological Aspects of X-Ray Clusters of Galaxies*, W.C. Seitter (ed), Kluwer Academic Publishers, 1994, p. 39
Geller M.J. & Beers T.C., 1982, PASP, 94, 421
Giovanelli R. & Haynes M.P., 1993, AJ, 105, 1271
Hasinger G., in *The X-ray background*, eds. Bacons & Fabian, C.U.P. 1992, p. 229
Henry J.P. & Briel U.G., 1995, ApJ, in print
Hickson P. 1982, ApJ, 255, 382
Hickson P., Mendes de Oliveira C., Huchra J.P., Palumbo G.G.C., 1992, ApJ, 399, 353
Hickson P., 1993, Astrophys. Lett. and Comm., 29, No. 1–3
Kindl E., 1990, PhD thesis, University of British Columbia
King I.R. 1962, AJ, 67, 471
Jones C. & Forman W., 1984, 276, 38
Mendes de Oliveira C., Gonzales J.J., Visvanathan N., Hickson P., 1994, in *The 9th IAP Astrophysics Meeting: Cosmic Fields*, F. Bouchet & M. Lachieze-Rey (eds), Ed. Frontieres, Gif-sur-Yvette, p563
Mendes de Oliveira C., Mackenty J., Hickson P., 1995, in preparation
O'Dea C.P. & Owen F.N., 1985, AJ, 90, 927
Owen F.N., Ledlow M.J., Keel W.C., 1995, AJ, 109, 14
Palumbo G.G.C, Saracco P., Hickson P., Mendes de Oliveira C., 1995, AJ, in press
Pildis R.A., Bregman J.N., Evrard A.E., 1995, ApJ, in press
Ponman T.J. & Bertram D., 1993, Nature, 363, 51
Postman M., Huchra J.P., Geller M.J., 1992, ApJ, 384, 404
Raymond J.C. & Smith B.W., 1977, ApJS, 35, 419
Ramella M., Diaferio A., Geller M.J., Huchra J.P., 1994, AJ, 107, 1623
Rangarajan F.V.N., White D.A., Ebeling H., Fabian A.C., 1995, submitted to MNRAS
Rhee G.F.R.N., van Haarlem M.P., Katgert P., 1991, A&A, 246, 301
Rood H.J. & Williams B.A., 1989, ApJ, 339, 772
Rood H.J. & Struble M.F., 1994, PASP, 106, 413
Schombert J.M., 1987, ApJS, 64, 643
Snowden S.L., Plucinsky P.P., Briel U., Hasinger G., Pfeffermann E., 1992, ApJ, 393, 819
Snowden S.L. & Freyberg M.J., 1993, ApJ, 404, 403
Snowden S.L., McCammon D., Burrows D.N., Mendenhall J.A., 1994, ApJ, 424, 714
Stark A.A., Gammie C.F., Wilson R.W., Bally J., Linke R.A., Heiles C., Hurwitz M., 1992 ApJS, 79, 77
Struble M.F. & Rood H.J. 1987, ApJS, 63, 555
Sulentic J.W., 1987, ApJ, 322, 605
Voges W., 1992, Proceedings of Satellite Symposium 3, ESA ISY-3, p9
White D.A., Fabian A.C., Johnstone R.M., Mushotzky R.F., Arnaud K.A., 1991, MNRAS, 252, 72
White D.A. & Fabian A.C., 1995, MNRAS, 273, 72
White D.A., 1995, in preparation
White S.D.M & Frenk C.S., 1991, ApJ, 379, 52
White S.D.M., Briel U.G., Henry J.P., 1993, MNRAS, 261, L8
Zabludoff A.I., Huchra J.P., Geller M.J., 1990, ApJS, 74, 1
Zucca E., Zamorani G., Scaramella R., Vettolani G., 1993, ApJ, 407, 470